# Security Improvement of an Image Encryption Based on mPixel-Chaotic-Shuffle and Pixel-Chaotic-Diffusion

**Musheer Ahmad**
*Department of Computer Engineering, Faculty of Engineering and Technology, Jamia Millia Islamia, New Delhi-110025, India*
E-mail: musheer.cse@gmail.com

**Hamed D. Alsharari**
*Department of Electrical Engineering, College of Engineering, AlJouf University, AlJouf, Kingdom of Saudia Arabia*
E-mail: hamed_100@yahoo.com

**Munazza Nizam**
*Department of Information Technology, Galgotia College of Engineering & Technology, Greater Noida, India*
E-mail: munazzanizam23@gmail.com

**Abstract**

In this paper, we propose to improve the security performance of a recently proposed color image encryption algorithm which is based on multi-chaotic systems. The existing cryptosystem employed a pixel-chaotic-shuffle mechanism to encrypt images, in which the generation of shuffling sequences are independent to the plain-image/cipher-image. As a result, it fails to the chosen-plaintext and known-plaintext attacks. Moreover, the statistical features of the cryptosystem are not up to the standard. Therefore, the security improvements are framed to make the above attacks infeasible and enhance the statistical features as well. It is achieved by modifying the pixel-chaotic-shuffle mechanism and adding a new pixel-chaotic-diffusion mechanism to it. The keys for diffusion of pixels are extracted from the same chaotic sequences generated in the previous stage. The simulation analyses and studies are performed to demonstrate that the updated version of cryptosystem has better statistical features and resistant to the chosen-plaintext and known-plaintext attacks than the existing algorithm.

**Keywords:** Multiple Chaotic Systems, Image Cryptosystem, Attacks, Modified Pixel-Chaotic-Shuffle, Pixel-Chaotic-Diffusion

## 1. Introduction

In today's world of technological advancements in web, multimedia and wireless networks, the multimedia data such as digital images, audio, video becomes a crucial means of communication. It leads to easiness in unauthorized access, illegal usage, malicious alteration and disruption of sensitive multimedia data for intruders and attackers. So, there is an increasing demand for building robust and efficient security methods for privacy protection of digital multimedia data while transmitting them over the Internet and wireless networks. A possible conventional solution is to use encryption techniques to secure sensitive data. Encryption is the mathematical transformation of plaintext data into unintelligible form to provide data confidentiality, integrity, authentication and non-repudiation. It provides an end-to-end secure communication. The core idea behind encryption is to protect plaintext data that can only be apprehensible on decryption with correct secret key by a legal recipient. A strong encryption technique should have high statistical features, resistant to cryptographic attacks and fulfill the classical Shannon requirement of confusion and diffusion [1]. Confusion makes the relationship between the key and ciphertext as complex as possible, where as diffusion reshuffle bits of plaintext so that any redundancy in the plaintext is spread out over the ciphertext [2]. There are a number of traditional encryption techniques like DES, AES, IDEA, RSA, etc [2-4]. These techniques are effective in providing good confusion and diffusion for encrypting text data, but these number-theory based encryption techniques are not well-suited for multimedia data. The multimedia data is usually very large-sized and bulky; its adjacent pixels/frames have high correlation, has spatial and temporal redundancies. Encrypting such data using traditional techniques incurs high computing power, large computation time and high expenses for real-time multimedia applications like video conferencing, image-surveillance, image-based military and satellite communication etc. Hence, it demands better solutions to resolve the security problems of multimedia data effectively.

The chaotic signals have several features that resembles to some cryptographic properties like (1) The ergodicity property of chaotic signals resembles the confusion property in cryptography, (2) High sensitivity of chaotic signals to their initial conditions/system-parameters resembles the diffusion property of cryptography and (3) Noise-like behavior of chaotic sequences resembles the key sequences used in cryptography. The highly sensitive response of chaotic systems to initial conditions makes their trajectory unpredictable and highly random. Moreover, the generation of discrete chaotic signals using chaotic systems often requires low cost. Consequently, the chaos-based cryptography has gained attention of researchers and academician to develop chaos-based methods for securing multimedia data. The first chaos based image encryption algorithm was proposed by R. Matthews [5]. Chaos-based cryptosystems provide strong encryption effect, better statistical features and high security. As a result, they are extensively exploited for encrypting multimedia images and videos [6-21]. However, some of them suffer from serious security flaws and are incompetent to withstand even the classical and other types of cryptographic attacks, as exposed by many cryptanalysts [22-28].

Most of the chaos-based image encryption algorithms are based on confusion and diffusion techniques. Confusion technique shuffles the positions of pixels in plain-image to get visually disordered and unrecognizable image. Diffusion technique alters the statistical characteristics of image by modifying the gray-values of pixels. Deployment of confusion and diffusion stages together provides higher encryption effect, robustness and security. Scharinger [6] suggested a chaotic kolmogorov-flow based image encryption algorithm. The image is permuted through a key-controlled system and gray-value substitution is based on a shift-registered pseudo-random number generator. Fridrich [7] suggested a cryptosystem in which the permutation of image pixels is done using 2D chaotic map. In the diffusion stage, the pixel values are changed depending on the accumulated effect of all previous pixels. Chen *et al.* [8] extended a 2D Cat map to 3D map to de-correlate relations among image pixels. The confusion is done by permuting the image pixels through extended 3D Cat map. Logistic map and 3D Chen's system are employed to diffuse the image. Mao *et al.* [9] extended the concept of Chen *et al.* [8] by employing a 3D Baker's map instead of 3D Cat map at the confusion stage. Fu *et al.* [10] employed preprocessed sequences generated by 3D Lorenz system to perform the confusion and diffusion. The encryption speed is improved as the position permutation and gray-value substitution of pixel is done in one iteration operation. Dongming *et al.* [11] constructed an ergodic matrix using Logistic map after executing an optimized preprocessing to permute the pixels of plain-image and a discretized Chen system is employed at diffusion stage to improve the encryption performance. Liu *et al.* [12] used an improved 3D Cat map for pixels shuffling and gray-value substitution, where the control parameters of Cat map are generated through 2D Henon's map and 2D coupled Logistic map is employed to generate parameters of substitution. Patidar *et al.* [13] presented a loss-less symmetric color image encryption algorithm based on chaotic 2D standard map and 1D Logistic map. In their algorithm, there are four rounds: two rounds for substitution/confusion and two rounds for diffusion. The first round of substitution/confusion is achieved with the help of intermediate XORing keys calculated from secret key. Then rounds for horizontal and vertical diffusions are completed by mixing the properties of horizontally and vertically adjacent pixels, respectively. In the last round, substitution/confusion is accomplished by generating an intermediate chaotic key stream image using chaotic standard and logistic maps. Tang *et al.* [14] suggested a new image encryption scheme using coupled map lattices (CML) with time-varying delays. A discretized tent map is employed to permute the positions of image pixels and a delayed CML is used to confuse the relationship between the plain-image and the cipher-image. The features of fourth-order hyper-chaotic system are improved and explored by Zhu [15] for designing an image encryption method. In [16], a new image encryption scheme based on coupling of chaotic function and XOR operator is provided. The scheme has the features of high security, sensitivity and randomness. Hongjun *et al.* [17] designed a stream-cipher algorithm based on one-time keys and robust chaotic maps in order to get high security and improved dynamical degradation, where the initial conditions are generated by the MD5 of mouse positions. This makes the algorithm robust against noise and makes known attacks infeasible. In [18], the authors proposed an image encryption algorithm by exploring the features of DNA computing and chaotic logistic function.

In 2009, Huang *et al.* [20] proposed pixel-chaotic-shuffle based color image encryption algorithm. The algorithm uses four three-dimensional chaotic systems for pixels bits shuffling. Solak *et al.* [22] breaks their scheme successfully by cracking the shuffling sequences that are equivalent keys of cryptosystem. This paper presents security improvements to make existing cryptosystem robust against Solak *et al.* attacks and to enhance its statistical features.

## 2. Proposed Security Improvements

The Huang *et al.* [20] proposed a pixel-chaotic-shuffle mechanism which utilizes four 3D chaotic systems namely the Henon map, the Lorenz map, the Chua map and the Rossler map for encrypting color images. The four 3D chaotic systems used in the design are described by Eqns (1)-(4) in Section 2.1 of Ref. [20]. We refer them as Eqns (1)-(4) in the later part of this paper. These chaotic systems are iterated and processed to generate the shuffling sequences. In pixel-chaotic-shuffle mechanism, the whole idea of encryption of RBG images involves two phases. In the first phase, the bits of binarized-image component are permuted vertically by performing column-wise indexing and shuffling. In the second phase, the 8-bits of each pixels of image component are rearranged horizontally within themselves through row-wise indexing and shuffling. One major shortcoming of Huang *et al.* algorithm is that the generation of shuffling sequences is independent to the pending plain-image or the cipher-image. As a consequence, it generates same sequences for encrypting different plain-images. Another reason which makes the work of attacker easier is that each color component of plain-image is processed separately and independently. These shortcomings facilitate the cryptanalysts Solak *et al.* [22] to break their algorithm.

We propose security improvements in Huang *et al.* algorithm with similar basic description, parameters and functions used. The improvements are framed to rule out the aforesaid shortcomings of the existing method. A modified pixel-chaotic-shuffle mechanism is presented to (1) create a dependency of twelve shuffling sequences to the plain-image to be encrypted and (2) process three components of color image collectively and dependently. Moreover, the modified pixel-chaotic-shuffle mechanism is appended by proposed pixel-chaotic-diffusion mechanism to enhance the statistical features of updated version. As a result, the improvements make the cryptanalysis, executed in [22], infeasible and also improves the statistical features of cryptosystem.

The plaintext color image $P$ of size $m \times n \times 3$, is first vectorized using raster-scan method (in $R \rightarrow G \rightarrow B$ order) to obtain an array of size $N \times 3$, where $N = mn$. The pixel's intensity values are decomposed into its binary equivalents of 8-bit format to form a binary image matrix $\xi$ of size $N \times 24$. To make the shuffling sequences dependent on plain-image, the total number of 1s in binarized color image $\xi$ is calculated, let it be $\Delta$. The four parameters $N_H$, $N_L$, $N_C$ and $N_R$ are evaluated based on the value of $\Delta$. The four chaotic systems with specified key parameters are iterated for $N_H$, $N_L$, $N_C$ and $N_R$ times and resulted chaotic values are discarded. It is done to achieve two purposes: (1) to establish a relation between the plain-image and the chaotic sequences or eventually the shuffling sequences and (2) to remove the transient effect of the chaotic systems used. The future trajectories of the four systems are solely controlled by the parameter $\Delta$, which is specific to the pending plain-image. Thus, it extracts information from the plain-image and utilizes it to iterate the chaotic systems. Consequently, an entirely different set of sequences are generated when encrypting a slightly different plain-image. It plays a key role in defeating the potential chosen-plaintext attack and known-plaintext attack. The 24-bits of each row of binary image matrix $\xi$ is manually arranged in a manner shown in Figure 1, to bring the initial confusion among $RGB$ pixels, let $\Psi_{rgb}$ be the matrix obtained. This way establishing the dependency of components on each other, this in turn increases the computation of cryptanalysis. Thus 8-bit pixel of each $R$, $G$, $B$ component, that was shuffled individually in [20], is replaced by 24-bit pattern for each $RGB$ pixel in the updated version. The procedure is then followed by column-wise indexing and shuffling, row-wise indexing and shuffling and pixel-chaotic-diffusion.

The following twelve chaotic sequences are obtained on applying next $mn$ iterations to each chaotic systems, $X_i(k)$, $Y_i(k)$ and $Z_i(k)$ where $i$ = 1, 2, 3, 4.

$$\begin{aligned} X_i &= \{x_i(1), x_i(2), \cdots\cdots, x_i(mn)\} \\ Y_i &= \{y_i(1), y_i(2), \cdots\cdots, y_i(mn)\} \\ Z_i &= \{z_i(1), z_i(2), \cdots\cdots, z_i(mn)\} \end{aligned}$$

To improve their stochasticity and randomness, these sequences are preprocessed using following formulation [10], where $k$ = 1, 2, ……, $mn$.

$$\begin{aligned} \hat{X}_i(k) &= \{X_i(k)\times 10^6 - floor(X_i(k)\times 10^6)\} \\ \hat{Y}_i(k) &= \{Y_i(k)\times 10^6 - floor(Y_i(k)\times 10^6)\} \\ \hat{Z}_i(k) &= \{Z_i(k)\times 10^6 - floor(Z_i(k)\times 10^6)\} \end{aligned} \tag{5}$$

Now, each member of the sequences lies in the interval of (0, 1). To quantify the randomness of above preprocessed sequences, they are transformed to binary sequences $bSeqX$, $bSeqY$ and $bSeqZ$ using a threshold of $\theta = 0.5$ by Eqn. (6)-(8). The standard NIST statistical test suite [29] is applied to evaluate the randomness performance of these sequences.

$$bSeqX_i(k) = \begin{cases} 0 & for \quad \hat{X}_i(k) < \theta \\ 1 & for \quad \hat{X}_i(k) \geq \theta \end{cases} \tag{6}$$

$$bSeqY_i(k) = \begin{cases} 0 & for \quad \hat{Y}_i(k) < \theta \\ 1 & for \quad \hat{Y}_i(k) \geq \theta \end{cases} \tag{7}$$

$$bSeqZ_i(k) = \begin{cases} 0 & for \quad \hat{Z}_i(k) < \theta \\ 1 & for \quad \hat{Z}_i(k) \geq \theta \end{cases} \tag{8}$$

The results of various statistical tests are listed in the Table 1. It is clear from the Table that all twelve sequences successfully passed the randomness tests as the associated *p_values* are higher than 0.01. These stochastically better preprocessed sequences are utilized to produce shuffling and encryption key sequences in proposed improved version.

**Table 1:** Randomness test results of twelve sequences by NIST statistical test suite

| **Statistical Test** | $bSeqX_1$ (*p_value*) | $bSeqX_2$ (*p_value*) | $bSeqX_3$ (*p_value*) | $bSeqX_4$ (*p_value*) | $bSeqY_1$ (*p_value*) | $bSeqY_2$ (*p_value*) | $bSeqY_3$ (*p_value*) | $bSeqY_4$ (*p_value*) | $bSeqZ_1$ (*p_value*) | $bSeqZ_2$ (*p_value*) | $bSeqZ_3$ (*p_value*) | $bSeqZ_4$ (*p_value*) | ***Results*** |
|---|---|---|---|---|---|---|---|---|---|---|---|---|---|
| *Frequency Test* | 0.838384 | 0.096299 | 0.350555 | 0.593732 | 0.838384 | 0.881516 | 0.962459 | 0.525157 | 0.844519 | 0.366986 | 0.832259 | 0.777631 | *All Success* |
| *Block Frequency Test* | 0.649814 | 0.334810 | 0.669590 | 0.918253 | 0.850038 | 0.354371 | 0.654242 | 0.497521 | 0.795345 | 0.803055 | 0.035018 | 0.401879 | *All Success* |
| *Cusum-Forward Test* | 0.694591 | 0.017580 | 0.208817 | 0.768087 | 0.698272 | 0.538884 | 0.611208 | 0.572753 | 0.701954 | 0.471638 | 0.979937 | 0.720381 | *All Success* |
| *Cusum-Reverse Test* | 0.877724 | 0.124895 | 0.447502 | 0.804052 | 0.880807 | 0.421359 | 0.654330 | 0.735115 | 0.877724 | 0.679896 | 0.912647 | 0.948732 | *All Success* |
| *Runs Test* | 0.124118 | 0.486930 | 0.277482 | 0.137872 | 0.124118 | 0.319077 | 0.712359 | 0.998736 | 0.126052 | 0.194433 | 0.566756 | 0.017443 | *All Success* |
| *Longest Runs Test* | 0.111325 | 0.103773 | 0.121785 | 0.080958 | 0.111325 | 0.580547 | 0.240356 | 0.676767 | 0.111325 | 0.188293 | 0.582790 | 0.507158 | *All Success* |
| *Rank Test* | 0.211605 | 0.621915 | 0.674257 | 0.014727 | 0.674257 | 0.918215 | 0.047407 | 0.669556 | 0.206921 | 0.776819 | 0.038096 | 0.472473 | *All Success* |
| *FFT Test* | 0.702631 | 0.296714 | 0.959403 | 1.000000 | 0.702631 | 0.779504 | 0.898737 | 0.721601 | 0.838657 | 0.460460 | 0.683845 | 0.524591 | *All Success* |
| *Linear Complexity Test* | 0.211212 | 0.846274 | 0.088353 | 0.191590 | 0.645386 | 0.912424 | 0.467533 | 0.165069 | 0.063691 | 0.846274 | 0.030934 | 0.543717 | *All Success* |
| *Serial Test* | 1.000000 | 1.000000 | 1.000000 | 1.000000 | 1.000000 | 1.000000 | 1.000000 | 1.000000 | 1.000000 | 1.000000 | 1.000000 | 1.000000 | *All Success* |

**Figure 1:** Schematic block diagram of the proposed improved algorithm.

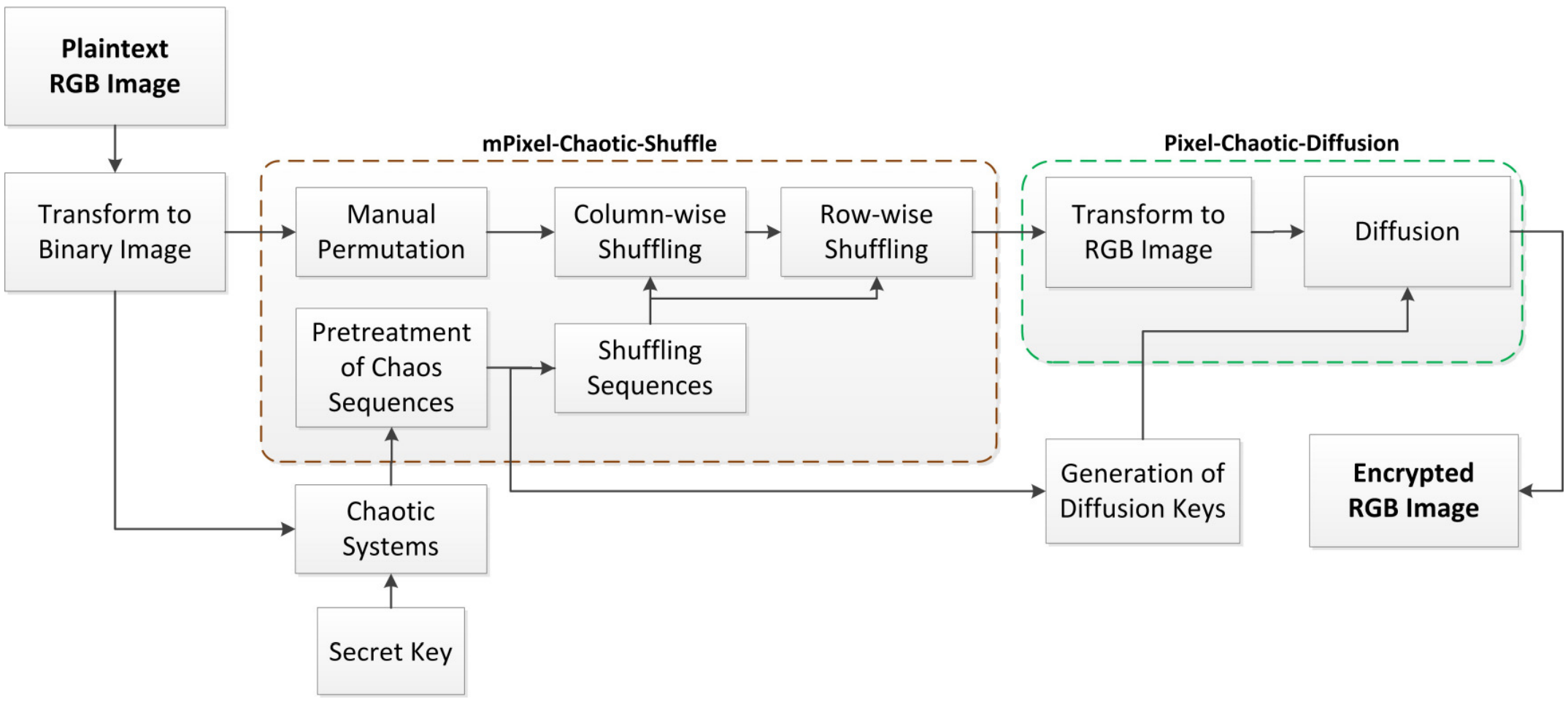

The block diagram of the proposed algorithm is depicted in Figure 1 and the algorithmic steps are as follows:

**Step 1**. Read the color image $P_{RGB}$ and prepare it to get binary image matrix $\xi_{rgb}$ of size $mn{\times}24$.

**[modified Pixel-Chaotic-Shuffle Stage]**

**Step 2**. Determine the number of 1's in the matrix $\xi_{rgb}$, let it be Δ and evaluate the parameters $N_H$, $N_L$, $N_C$ and $N_R$ from Δ as:

$$N_H = (\Delta) mod(997) + 829$$

$$N_L = (\Delta) mod(937) + 529$$

$$N_C = (\Delta) mod(1097) + 719$$

$$N_R = (\Delta) mod(397) + 1123$$

**Step 3**. Manually arrange the bits of image matrix $\xi_{rgb}$, as shown in Figure 2, to form matrix $\Psi_{rgb(mn\times 24)}$.

**Figure 2:** Manual arrangement of 24-bits of image matrix.

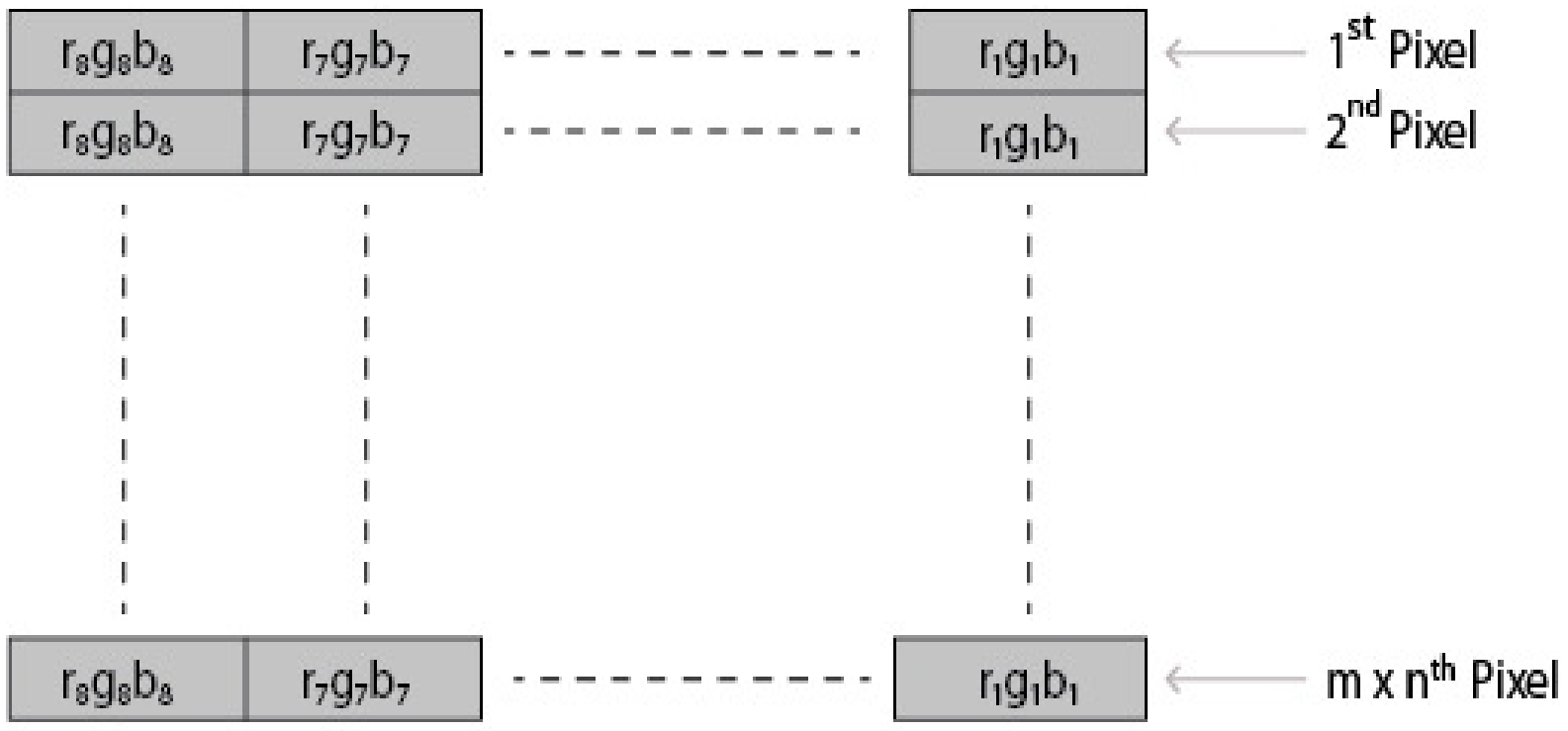

**Step 4**. Select the initial conditions and parameters for the four chaotic systems.

**Step 5**. Iterate the four chaotic systems of Eqns (1)-(4) for $N_H$, $N_L$, $N_C$ and $N_R$ times and discard the chaotic values.

**Step 6**. Again iterate the chaotic systems of Eqns (1)-(4) for next $mn$ times to capture the sequences $X_{1(\mu,1)}$ *to* $X_{4(\mu,1)}$, $Y_{1(\mu,1)}$ *to* $Y_{4(\mu,1)}$ and $Z_{1(\mu,1)}$ *to* $Z_{4(\mu,1)}$, where $\mu = 1, 2, 3, \ldots, mn$.

**Step 7.** Preprocess the chaotic sequences through Eqn. (5).

**Step 8**. Extract the indexing sequences $F_{X1} \sim F_{X4}$, $F_{Y1} \sim F_{Y4}$ and $F_{Z1} \sim F_{Z4}$ from preprocessed chaotic sequences $\hat{X}_1 \sim \hat{X}_4$, $\hat{Y}_1 \sim \hat{Y}_4$ and $\hat{Z}_1 \sim \hat{Z}_4$ as:

$$F_{X1} = sort\,(\hat{X}_{1\,(\mu,1)}) \quad F_{X2} = sort\,(\hat{X}_{2\,(\mu,1)}) \quad F_{X3} = sort\,(\hat{X}_{3\,(\mu,1)}) \quad F_{X4} = sort\,(\hat{X}_{4\,(\mu,1)})$$

$$F_{Y1} = sort\,(\hat{Y}_{1\,(\mu,1)}) \quad F_{Y2} = sort\,(\hat{Y}_{2\,(\mu,1)}) \quad F_{Y3} = sort\,(\hat{Y}_{3\,(\mu,1)}) \quad F_{Y4} = sort\,(\hat{Y}_{4\,(\mu,1)})$$

$$F_{Z1} = sort\,(\hat{Z}_{1\,(\mu,1)}) \quad F_{Z2} = sort\,(\hat{Z}_{2\,(\mu,1)}) \quad F_{Z3} = sort\,(\hat{Z}_{3\,(\mu,1)}) \quad F_{Z4} = sort\,(\hat{Z}_{4\,(\mu,1)})$$

where *sort(.)* is a sequencing index function defined in [20].

**Step 9**. Apply shuffle function $sq(\cdot)$ on $\Psi_{rgb}$ for column-wise shuffling as shown in Figure 3 using shuffling indices $F_{X1} \sim F_{X4}$, $F_{Y1} \sim F_{Y4}$ and $F_{Z1} \sim F_{Z4}$. The function $sq(\cdot)$ shuffles the bits of matrix $\Psi_{rgb}$ using above indexing sequences. Thus, we get a partially encrypted column shuffled binary image matrix as:

$$\psi_{ergb\mu} = [\psi_{ergb\mu 1}\psi_{ergb\mu 2}\cdots\cdots\cdots\psi_{ergb\mu 24}]$$

where,

$$\psi \mathrm{ergb}_{\mu i} = \begin{Bmatrix} sq(\psi \mathrm{rgb}_{\mu i}, Fx_1), i = 1,2 \\ sq(\psi \mathrm{rgb}_{\mu i}, Fx_2), i = 3,4 \\ sq(\psi \mathrm{rgb}_{\mu i}, Fx_3), i = 5,6 \\ sq(\psi \mathrm{rgb}_{\mu i}, Fx_4), i = 7,8 \\ sq(\psi \mathrm{rgb}_{\mu i}, Fy_1), i = 9,10 \\ sq(\psi \mathrm{rgb}_{\mu i}, Fy_2), i = 11,12 \\ sq(\psi \mathrm{rgb}_{\mu i}, Fy_3), i = 13,14 \\ sq(\psi \mathrm{rgb}_{\mu i}, Fy_4), i = 15,16 \\ sq(\psi \mathrm{rgb}_{\mu i}, Fz_1), i = 17,18 \\ sq(\psi \mathrm{rgb}_{\mu i}, Fz_2), i = 19,20 \\ sq(\psi \mathrm{rgb}_{\mu i}, Fz_3), i = 21,22 \\ sq(\psi \mathrm{rgb}_{\mu i}, Fz_4), i = 23,24 \end{Bmatrix}$$

and $\Psi_{rgb\mu i}$ is the $i$th bit of the $\mu$th row of matrix $\Psi_{rgb}$.

**Step 10**. Perform row-wise shuffling of bits within each row of matrix $\Psi_{ergb}$ obtained in above step in pairs of 2 using indices obtained in Step 7. Let the matrix obtained be $\Psi_{srgb}$ of size $mn \times 24$.

**Step 11**. Prepare the shuffled binary image matrix $\Psi_{srgb}$ to get RGB shuffled image $S_{RGB}$ of size $m \times n \times 3$.

**[Pixel-Chaotic-Diffusion Stage]**

**Step 12**. Decompose the shuffled image $S_{RGB}$ into three gray-scale images of $S_R$ (red), $S_G$ (green) and $S_B$ (blue), arrange their pixels in raster-scan order to get three 1D sequences as:

$$S_R = \{S_R(1), S_R(2), \cdots\cdots, S_R(mn)\}$$
$$S_G = \{S_G(1), S_G(2), \cdots\cdots, S_G(mn)\}$$
$$S_B = \{S_B(1), S_B(2), \cdots\cdots, S_B(mn)\}$$

**Step 13**. Extract the key sequences for diffusion using the preprocessed chaotic sequences $\hat{X}_1 \sim \hat{X}_4$, $\hat{Y}_1 \sim \hat{Y}_4$ and $\hat{Z}_1 \sim \hat{Z}_4$ obtained earlier ($k = 1 \sim mn$).

$$keyX_i(k) = \{floor(\hat{X}_i(k) \times 10^{14})\}\,\mathrm{mod}(256)$$
$$keyY_i(k) = \{floor(\hat{Y}_i(k) \times 10^{14})\}\,\mathrm{mod}(256)$$
$$keyZ_i(k) = \{floor(\hat{Z}_i(k) \times 10^{14})\}\,\mathrm{mod}(256)$$

**Step 14**. Choose $C_R(0)$, $C_G(0)$ and $C_B(0)$.

**Step 15**. Iterate the following for $j = 1 \sim mn$.

$t_1 = (C_B(j\text{-}1))mod(12);$ $t_2 = (C_R(j\text{-}1))mod(12);$ $t_3 = (C_G(j\text{-}1))mod(12);$

$$C_R(j) = S_R(j) \oplus \{C_R(j\text{-}1) + Key(j, t_1)\}mod(256)$$
$$C_G(j) = S_G(j) \oplus \{C_G(j\text{-}1) + Key(j, t_2)\}mod(256)$$
$$C_B(j) = S_B(j) \oplus \{C_B(j\text{-}1) + Key(j, t_3)\}mod(256)$$

the definition of *Key* is provided in Table 2.

**Step 16**. Combine the encrypted gray-scale images $C_R$, $C_G$ and $C_B$ to encrypted color image $C_{RGB}$ of size $m \times n \times 3$.

**Figure 3:** Column-wise shuffling of bits in vertical direction in matrix $\Psi_{rgb}$.

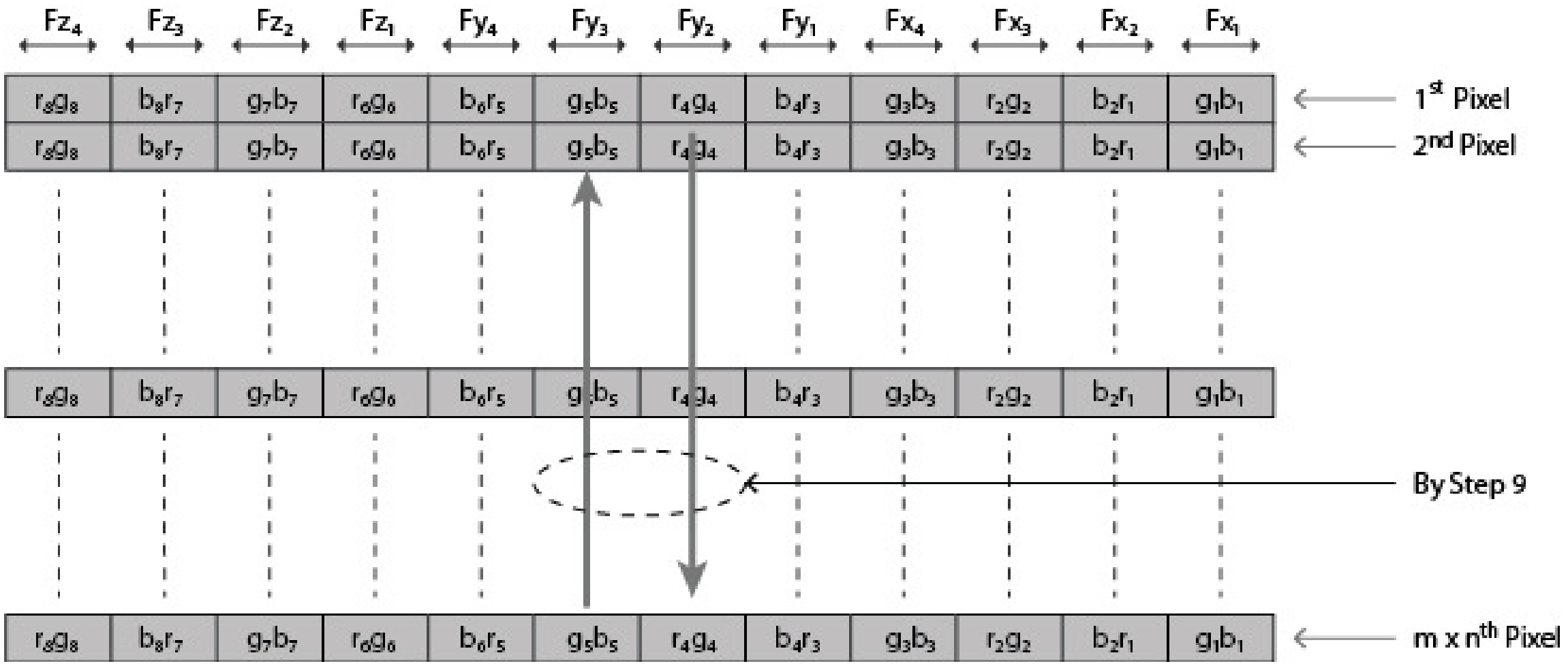

**Table 2:** Actual diffusion keys

| ***Key(j, t)*** | *keyX₁(j)* | *keyY₁(j)* | *keyZ₁(j)* | *keyX₂(j)* | *keyY₂(j)* | *keyZ₂(j)* |
|---|---|---|---|---|---|---|
| ***t*** | 0 | 1 | 2 | 3 | 4 | 5 |
| ***Key(j, t)*** | *keyX₃(j)* | *keyY₃(j)* | *keyZ₃(j)* | *keyX₄(j)* | *keyY₄(j)* | *keyZ₄(j)* |
| ***t*** | 6 | 7 | 8 | 9 | 10 | 11 |

## 3. Experimental Analyses and Results

Same standard color image *Lena* of size 256×256×3 is taken as test image to justify the improved security and robustness performance of proposed version. The two algorithms under consideration are implemented in MATLAB. The following simulation analyses are carried out to evaluate the security performance of both the algorithms.

### 3.1. Histogram Analysis

Image encryption performance evaluation via histograms is an effective criterion. Image histogram illustrates how pixels in an image are distributed. The histograms of *R, G, B* components of original *Lena* image and its encrypted image using existing algorithm are shown in Figure 4. Figure 5 depicts the histograms of *R, G, B* components of encrypted image obtained with proposed version. The histograms obtained for the case of existing technique has more number of peaks as compared to the histograms in proposed technique. The image with flat histogram level is analogous to a noise-image. The histograms shown in Figure 5 resemble that of a noisy image. It can be observed that more flat and fairly uniform histograms are obtained with proposed updated algorithm. The encrypted image shown in Figure 5 has cryptographically better pixels distribution than the pixels of encrypted image obtained with the existing Huang *et al.* algorithm.

**Figure 4:** Plain-image of '*Lena*' and histograms of its *R, G*, *B* components.

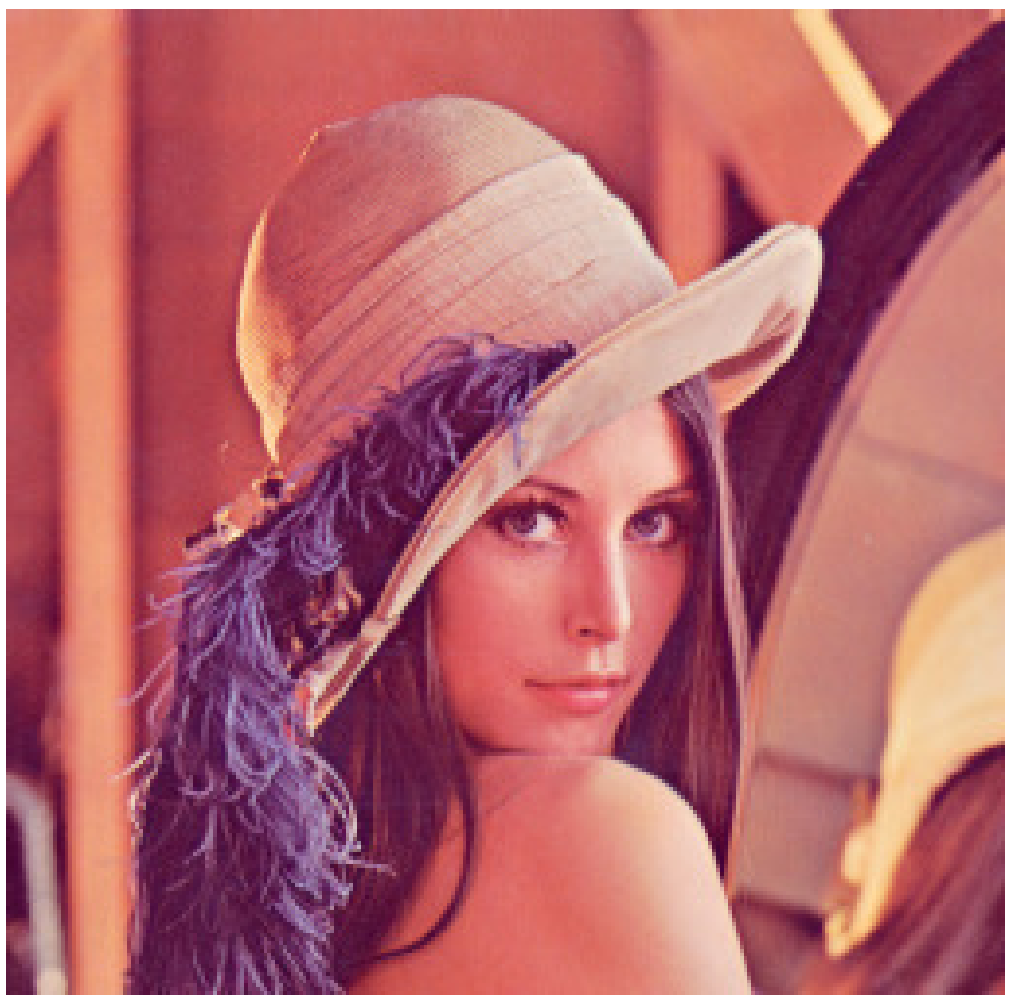

(a) Test image '*Lena*'

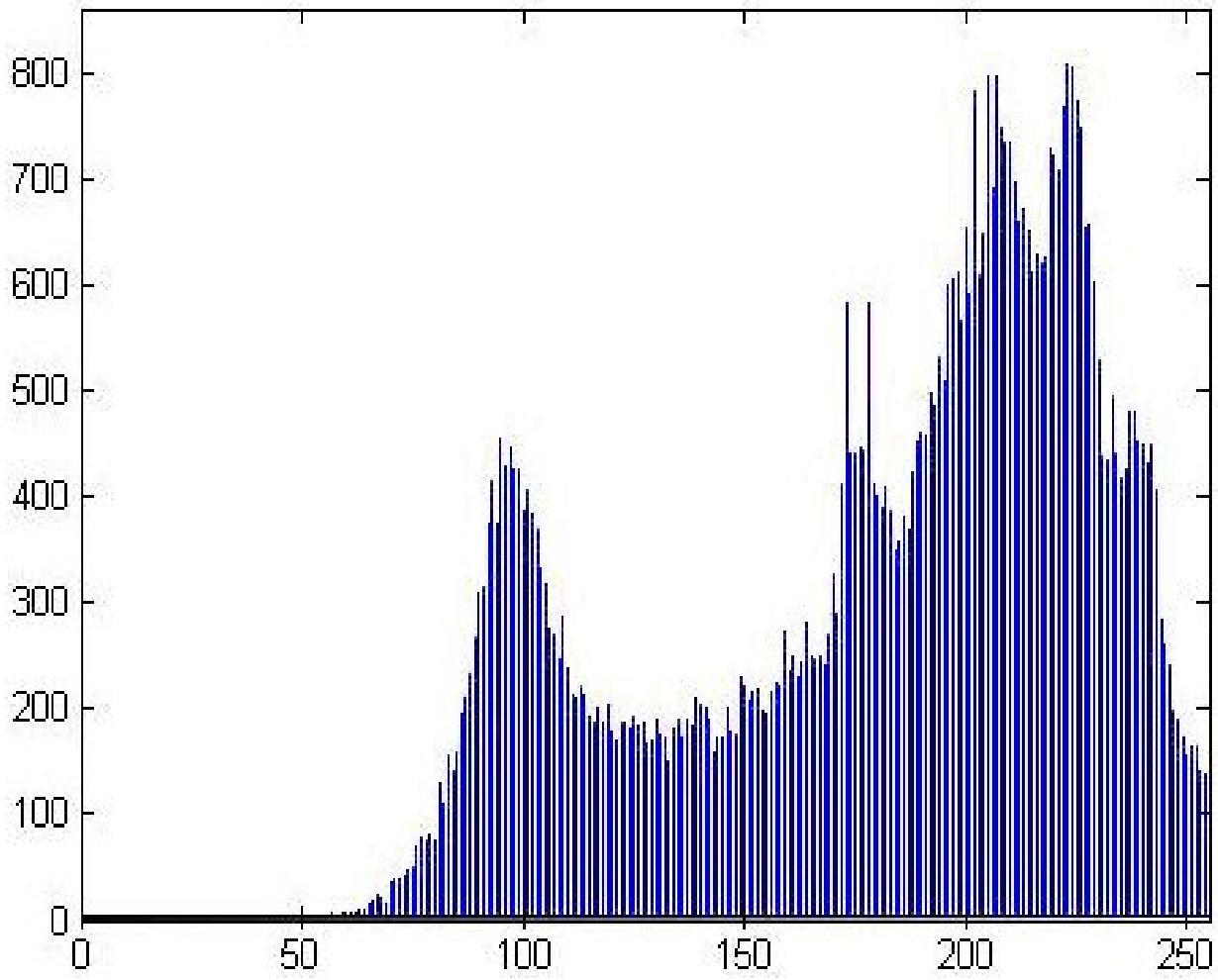

(b) *R* component

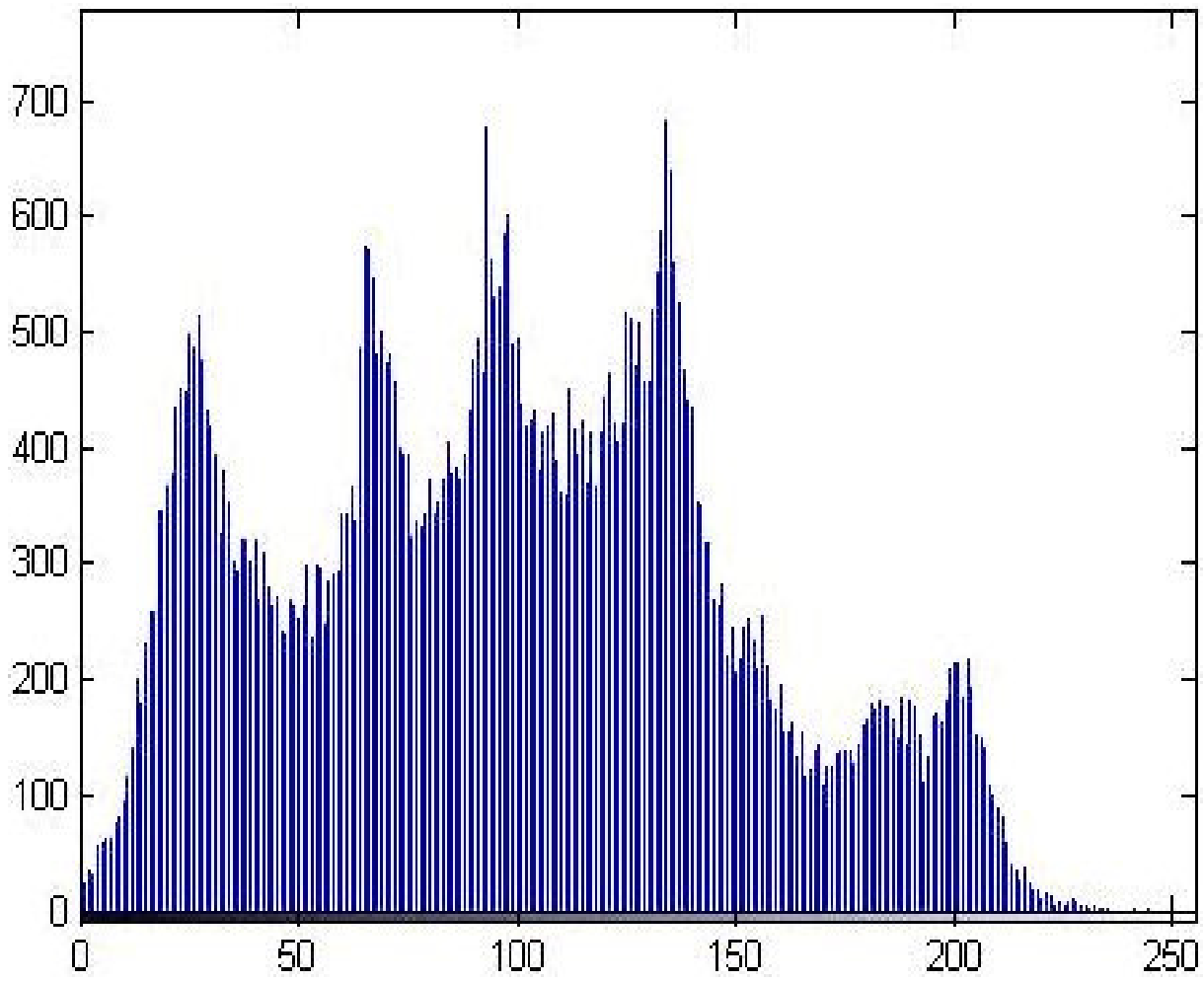

(c) *G* component

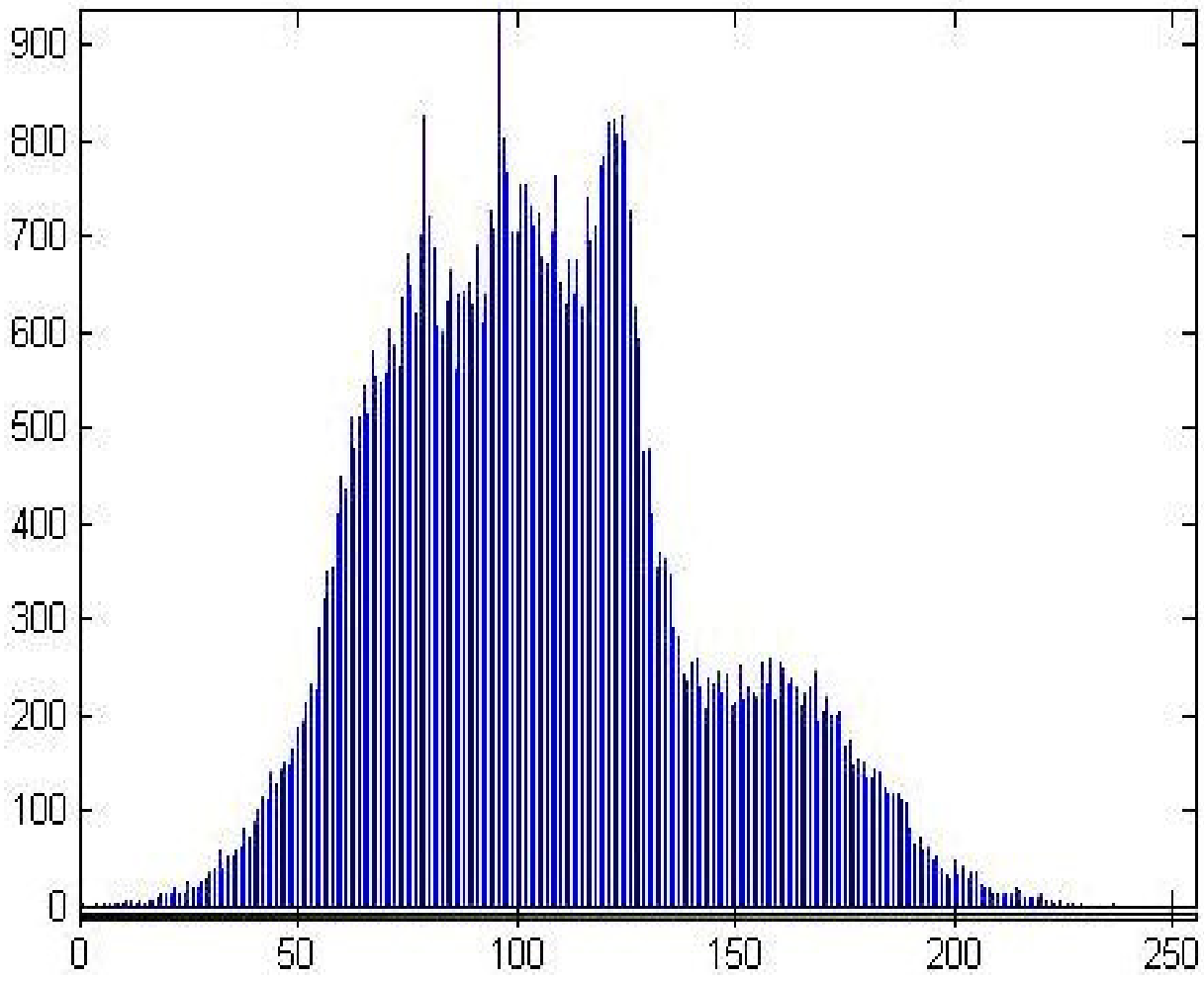

(d) *B* component

### 3.2. Mean Gray Value Analysis

In the proposed version, the statistical properties of color plain-images are improved in such a manner that encrypted images have good balance property. To quantify the balance property of images, the mean gray values of plain-image and encrypted images are evaluated and listed in Table 3. As can be seen from scores that no matter how gray-values of plain-image are distributed, the mean gray-values of encrypted images come out closer to 127.5 (ideal value for a gray-scale perfect noise image) as compared to the existing cryptosystem. This shows that the improved version doesn't provide any information regarding the distribution of gray values to the attacker in the encrypted images.

**Table 3:** Mean gray-values of images

| | **Red** | **Green** | **Blue** |
|---|---|---|---|
| Original | 180.22 | 99.05 | 105.41 |
| Huang *et al.* [20] | 147.78 | 109.03 | 127.51 |
| Xiao *et al.* [21] | NA | NA | NA |
| Proposed version | 127.51 | 127.97 | 127.93 |

### 3.3. Chi-Square Analysis

The security performance of an encryption method is also quantified through chi-square test [30]. It is a statistical test used to examine the variations of data from the expected value. The chi-square parameter $\chi 2$ is defined as:

$$\chi^2 = \sum_{i=1}^{256} \frac{(P_i - C_i)^2}{C_i} \tag{9}$$

Where $i$ is number of gray values, $P_i$ and $C_i$ are observed and expected occurrence of each gray value (0 to 255), respectively. The less the value of chi-square $\chi^2$ better will be the encryption performance of the scheme. The values of chi-square for images under study are listed in Table 4. It can be observed that chi-square values for proposed method are extremely low as compared to the values obtained for the original image and Huang *et al.* encrypted image. The extremely low values of chi-square validate that the proposed method offers fairly high encryption effect.

**Figure 4:** Encrypted image of '*Lena*' and histograms of its *R, G*, *B* components.

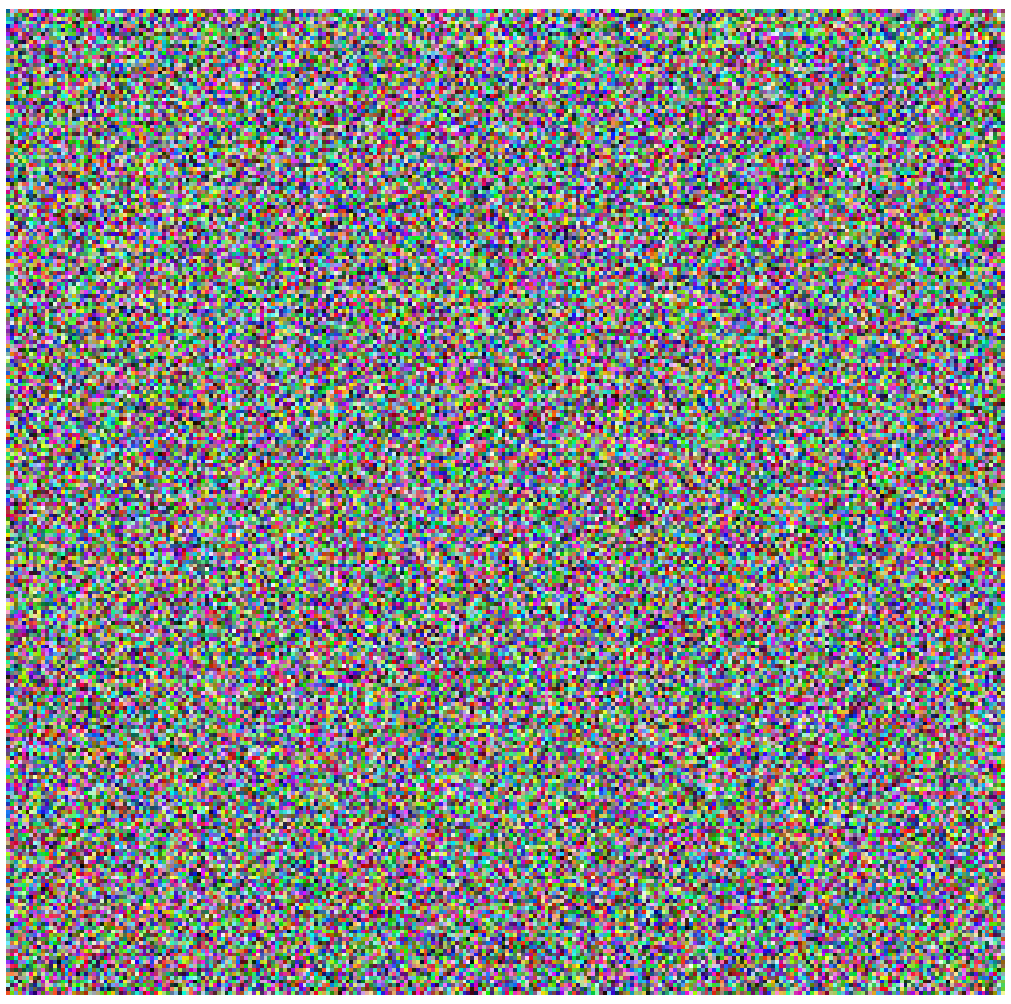

(c) Encrypted image

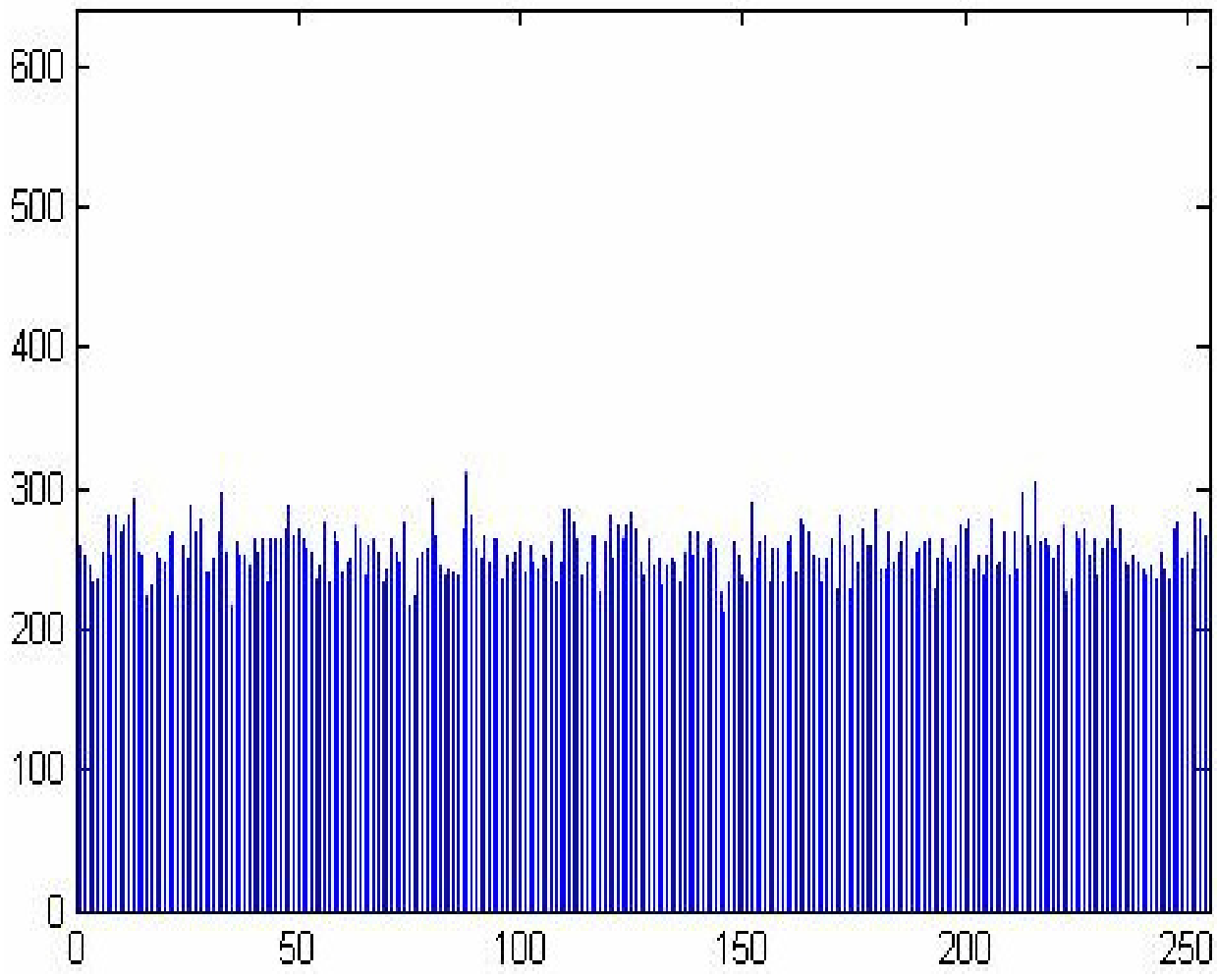

(d) *R* component

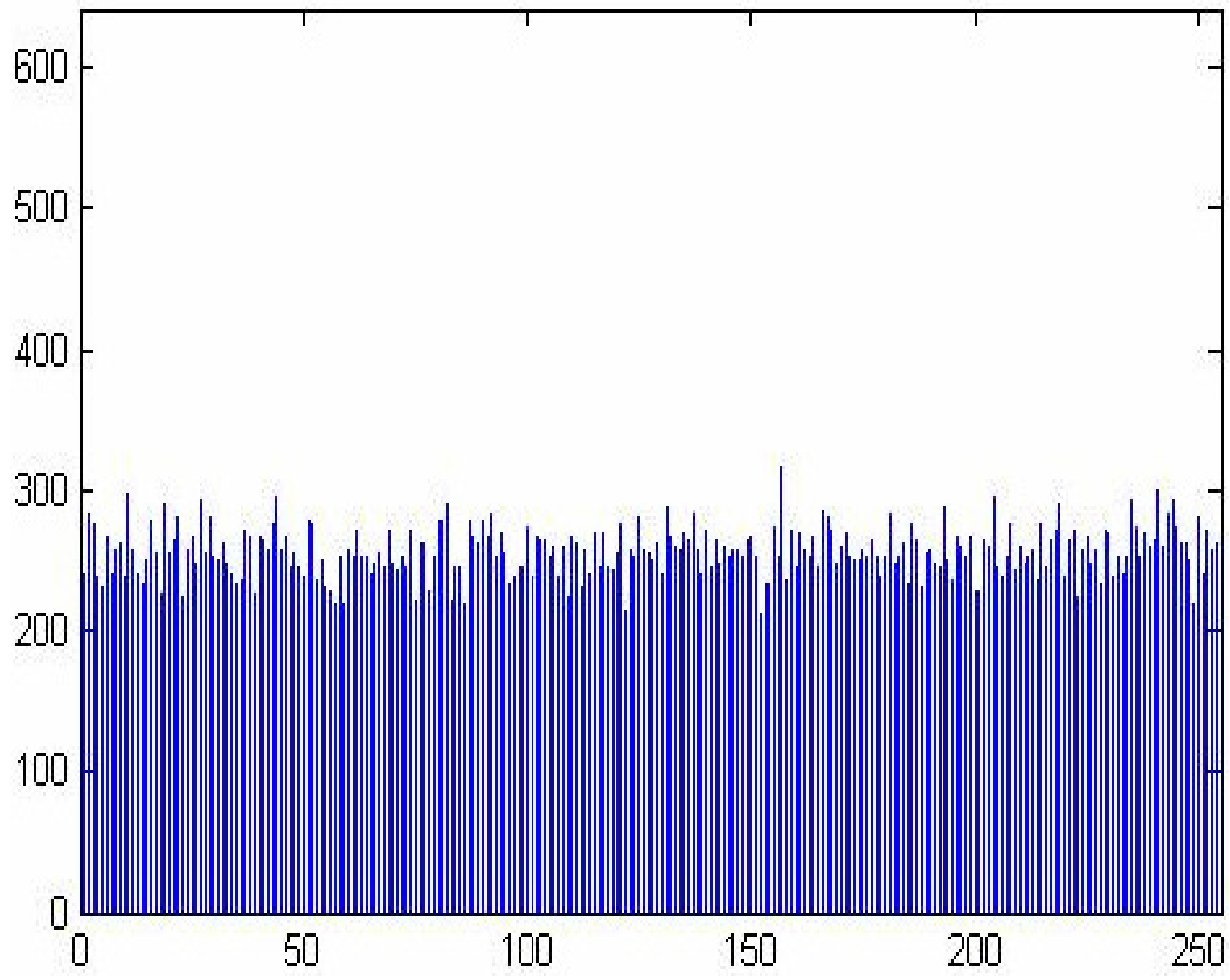

(c) *G* component

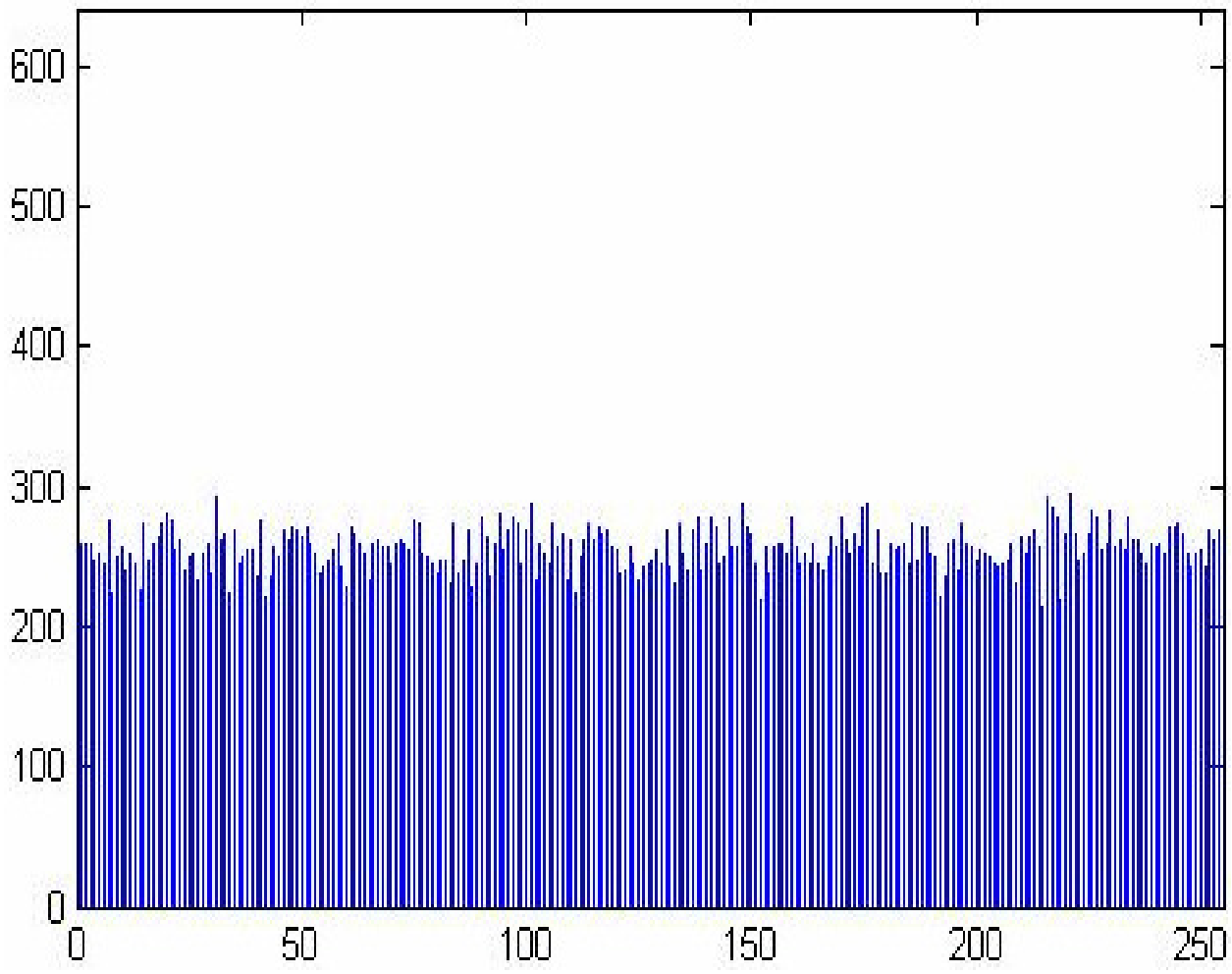

(d) *B* component

**Table 4:** Chi-square values of images

| | **Red** | **Green** | **Blue** |
|---|---|---|---|
| Original | 65274.12 | 30609.43 | 91931.33 |
| Huang *et al.* [20] | 12343.32 | 8570.18 | 37977.49 |
| Xiao *et al.* [21] | NA | NA | NA |
| Proposed version | 272.30 | 307.46 | 218.93 |

## 3.4. Correlation Analysis

The correlation between adjacent pixels of encrypted image should be as low as possible. For evaluating the correlation between the pixels in cipher image we randomly select pairs of adjacent pixels in image. The correlation coefficient is calculated as [19].

$$\rho = \frac{N\sum_{i=1}^{N}(x_i \times y_i) - \sum_{i=1}^{N} x_i \times \sum_{i=1}^{N} y_i}{\sqrt{(N\sum_{i=1}^{N} x_i^2 - (\sum_{i=1}^{N} x_i)^2) \times (N\sum_{i=1}^{N} y_i^2 - (\sum_{i=1}^{N} y_i)^2)}} \quad (10)$$

Where $x$ and $y$ are gray values of adjacent pixels and $N$ is the total number of pairs of pixels of an image. The values of correlation coefficients for the proposed and existing algorithm are given in Table 5. The proposed algorithm provides lower value of $\rho$ as compared to existing algorithms, thus our algorithm outperforms both Huang *et al.* [20] and Xiao *et al.* [21] algorithms.

**Table 5:** Correlation coefficient of pixels in images

| | **Horizontal** | **Vertical** | **Diagonal** |
|---|---|---|---|
| Original | 0.9597 | 0.9792 | 0.9570 |
| Huang *et al.* [20] | 0.1257 | 0.0581 | 0.0226 |
| Xiao *et al.* [21] | 0.0631 | 0.0226 | -0.0192 |
| Proposed version | -0.00513 | 0.00339 | -0.00373 |

## 3.5. Entropy Analysis

Information entropy of an image is a basic criterion used to depict the randomness of data. A greater value of information entropy shows a more uniform distribution of gray values of image. The entropy $H$ of a message source $M$ can be computed as:

$$H(M) = \sum_{i=0}^{255} p(m_i) \log\left(\frac{1}{p(m_i)}\right) \quad (11)$$

Where $p(m_i)$ represents the probability of symbol $m_i$ and the entropy is expressed in bits. If the source $M = \{m_0, m_1, \ldots\ldots, m_{255}\}$ emits $2^8$ symbols with equal probability, then the entropy $H(M) = 8$, which corresponds to a true-random source and represents the ideal value of entropy for message source. It is clear that the entropy scores for proposed algorithm are higher and closer to the ideal value than those computed with existing algorithm.

**Table 6:** Information entropy of images

| | **Red** | **Green** | **Blue** |
|---|---|---|---|
| Original | 7.2359 | 7.5689 | 6.9179 |
| Huang *et al.* [20] | 7.8501 | 7.9028 | 7.5582 |
| Xiao *et al.* [21] | NA | NA | NA |
| Proposed version | 7.9970 | 7.9966 | 7.9976 |

### 3.6. NPCR and UACI Analysis

The NPCR and UACI are two most significant quantities that quantify the strength of encryption algorithms. NPCR is the measure of absolute number of pixels change rate and UACI computes average difference of color intensities between two images when the change in one image is subtle. The NPCR and UACI values can be evaluated by Eqns. (12) and (13), where $T$ denotes the largest supported gray-value compatible with image format, $|\cdot|$ denotes the absolute value function [31].

$$\text{NPCR: } N(C_1, C_2) = \sum_{ij} \frac{D(i,j)}{m \times n} \times 100\% \tag{12}$$

$$\text{UACI: } U(C_1, C_2) = \frac{1}{m \times n} \sum_{ij} \frac{|C_1(i,j) - C_2(i,j)|}{T} \times 100\% \tag{13}$$

$$D(i,j) = \begin{cases} 0 & if \quad C_1(i,j) \neq C_2(i,j) \\ 1 & if \quad C_1(i,j) = C_2(i,j) \end{cases}$$

A pixel of plain-image $P_1$ is randomly chosen and is set to 0, let this new image be named $P_2$. Let $C_1$ and $C_2$ be the cipher images of images $P_1$ and $P_2$. NPCR and UACI values between $C_1$ and $C_2$ are calculated for the two schemes and listed in Table 7. Sufficiently high NPCR/UACI scores for $C_1$ and $C_2$ are usually considered as strong resistance to differential attacks. The Table shows that a tiny change in the plain image results almost no change for existing cryptosystem. However, it causes a significantly large difference in proposed method i.e. the proposed version is highly sensitive to a small change in the plain-image.

**Table 7:** NPCR and UACI between $C_1$ & $C_2$

| | **NPCR** | | | **UACI** | | |
|---|---|---|---|---|---|---|
| | Red | Green | Blue | Red | Green | Blue |
| Huang *et al.* [20] | 0.0045 | 0.0030 | 0.0030 | 0.00046 | 0.00114 | 0.00010 |
| Xiao *et al.* [21] | NA | NA | NA | NA | NA | NA |
| Proposed version | 99.596 | 99.637 | 99.577 | 33.335 | 33.467 | 33.563 |

Now, the NPCR and UACI between $P_1$ and $C_1$ are evaluated and listed in Table 8. The scores determine the deviation of encrypted image from its plain-image. It is evident from the comparison that the NPCR values are comparable and UACI scores are significantly better than the scores obtained with existing two algorithms.

**Table 8:** UACI between $P_1$ & $C_1$

| | **NPCR** | | | **UACI** | | |
|---|---|---|---|---|---|---|
| | Red | Red | Green | Blue | Green | Blue |
| Huang *et al.* [20] | 99.42 | 99.60 | 99.54 | 24.94 | 27.66 | 24.94 |
| Xiao *et al.* [21] | 99.54 | 99.55 | 99.46 | 23.91 | 27.18 | 23.91 |
| Proposed version | 99.63 | 99.57 | 99.53 | 27.56 | 30.67 | 27.56 |

### 3.6. Resistance to CPA/KPA Attacks

In the proposed version, the generation of shuffling sequences is made dependent to the pending image information in such a way that a tiny different plain-image results in distinct shuffling sequences, which in turn produce totally different encrypted image. Moreover, the components of the pending image are processed collectively and dependently. These improvements make the attacks executed in [22] infeasible and impossible. So, proposed updated version can resist the chosen-plaintext and known-plaintext attacks.

## 4. Conclusion

In this paper an updated version of color image encryption algorithm has been proposed. The shortcomings of existing technique are eliminated by dynamically changing the shuffling sequences whenever there is a tiny change in plain-image. It is achieved by extracting information specific to the pending plain-image and using it to generate the shuffling sequences. The *R, G, B* components of image are operated collectively and dependently. This guarantees the robustness of the proposed algorithm against CPA/KPA attacks. The statistical features of the algorithm are further bettered by adding a pixel-chaotic-diffusion stage to it, where the diffusion keys are obtained out of the chaotic sequences generated earlier. The NPCR and UACI scores show that proposed version is very sensitive to a slight change in the plain-image. Several other simulation analyses and comparative studies validate the improved security performance of the proposed version.